\documentclass[sigconf]{acmart}

\usepackage[T1]{fontenc}
\usepackage{graphicx}
\usepackage{stfloats}
\usepackage{tikz}
\usetikzlibrary{calc,positioning,decorations.pathreplacing,arrows.meta,fit}
\usepackage{adjustbox}
\usepackage[most]{tcolorbox}
\usepackage{listings}
\usepackage{multirow}
\usepackage{booktabs} 
\usepackage{xurl}
\usepackage{hyperref}
\usepackage{enumitem}

\tcbset{
  question/.style={
    enhanced,
    frame hidden,
    colback=white,
    boxrule=0pt,
    sharp corners,
    overlay={
      \draw[line width=0.5pt] (frame.south west) -- ++(0,0.3);
      \draw[line width=0.5pt] (frame.south west) -- ++(0.3,0);
      \draw[line width=0.5pt] (frame.north west) -- ++(0,-0.3);
      \draw[line width=0.5pt] (frame.north west) -- ++(0.3,0);
      \draw[line width=0.5pt] (frame.south east) -- ++(0,0.3);
      \draw[line width=0.5pt] (frame.south east) -- ++(-0.3,0);
      \draw[line width=0.5pt] (frame.north east) -- ++(0,-0.3);
      \draw[line width=0.5pt] (frame.north east) -- ++(-0.3,0);
    }
  }
}

\tcbset{
  answer/.style={
    colback=gray!7,
    colframe=gray!7,
    boxrule=0pt,
    left=5pt,
    right=5pt,
    top=5pt,
    bottom=5pt
  }
}

\AtBeginDocument{%
  }

\setcopyright{acmlicensed}
\copyrightyear{2026}
\acmYear{2026}
\acmConference[ASE '26]{IEEE/ACM Automated Software Engineering (ASE) Conference}{October 12--16,
  2026}{Munich, Germany}
\begin{document}

\title{Transformers for Program Termination}

\author{Yoav Alon}
\authornote{Both authors contributed equally to this research.}
\email{yoav.alon@bristol.ac.uk}
\affiliation{%
  \institution{University of Bristol}
  \country{UK}
}

\author{Cristina David}
\authornotemark[1]
\email{cristina.david@bristol.ac.uk}
\affiliation{%
  \institution{University of Bristol}
  \country{UK}
}

\renewcommand{\shortauthors}{Alon et al.}

\begin{abstract}
Determining whether a program terminates is a core challenge in program analysis with direct implications for correctness, verification, and security. We investigate whether transformer architectures can recognise termination patterns directly from source code and how their strengths can be amplified through ensembles. 
To overcome the extreme scarcity of non-terminating examples, we design an ensemble framework of compact transformer encoders, systematically trained with a suite of imbalance-aware loss functions and class-aware sampling techniques.
By combining models trained with distinct loss functions, our ensembles achieve substantially stronger performance than any single transformer, \emph{outperforming both powerful off-the-shelf LLMs and graph-based methods}. Finally, we introduce an attribution pipeline that produces syntax-aware explanations for the termination estimation.
\end{abstract}

\ccsdesc[500]{Computing methodologies~Artificial intelligence}
\ccsdesc[300]{Computing methodologies~Knowledge representation and reasoning}
\ccsdesc[500]{Computing methodologies~Neural networks}

\keywords{Transformers, Program Termination}

\received{26 March 2026}
\maketitle

\section{Introduction}

Determining whether a program eventually terminates is a central problem in program analysis, with direct implications for correctness, verification, and security. Non-terminating behaviour can lead to denial-of-service vulnerabilities, memory exhaustion, or system deadlocks, making automated termination detection a long-standing research goal. While formal methods compute ranking functions and recurrence sets to provide rigorous guarantees, they remain incomplete in general~\cite{DBLP:conf/pldi/CookPR06, DBLP:conf/tacas/CookSZ13, DBLP:conf/esop/DavidKL15, DBLP:conf/pldi/ChenHLLTTZ18}, and typically target specific classes of programs or abstractions.

As a complementary strand, shifting the objective from formal verification to empirical prediction, a growing body of work employs neural approaches to estimate program termination. Early work by Giacobbe et al.\ introduced Neural Termination Analysis~\cite{neural-termination}, where neural networks are trained as ranking functions from sampled program traces and then verified using SMT solving. While this line of work maintains formal guarantees, it remains limited to settings where termination certificates can be verified symbolically. Building on purely neural methods, Alon and David proposed classifying termination directly from code graphs using Graph Neural Networks (GCNs and GATs)~\cite{TerminationGNN}, exploring graph-based representations of programs. Replicating and extending this approach, Liu et al.\ presented TerGEC~\cite{TerGEC}, a graph-enhanced contrastive framework that combines intra- and inter-class semantics with imbalance-aware training objectives.

Given the rapid progress of transformer models in recent years, a natural question arises: \emph{can transformer architectures learn to recognise termination patterns directly from source code?} Unlike graph-based approaches, transformers capture long-range dependencies and token-level interactions without requiring handcrafted program representations or auxiliary program graphs. If termination cues are reflected in the syntax and structure of code---such as loop guards, recursive conditions, or counter updates---then transformers should, in principle, be able to discover them.

Very recent work has begun to explore this question at scale. Sultan et al.\ investigate termination prediction using large language models (LLMs)~\cite{DBLP:journals/corr/abs-2601-18987}, evaluating frontier systems such as GPT-5 and Claude Sonnet on the SV-Comp dataset~\cite{sv-comp}. Their results show that sufficiently large models can achieve performance competitive with specialised verification tools. However, these models are orders of magnitude larger than typical open-weight or locally deployable systems, requiring substantial computational resources and access to external infrastructure.

In this paper, we study transformer-based termination prediction through the lens of four practical challenges: \emph{deployability}, \emph{task adaptation}, \emph{extreme class imbalance}, and \emph{interpretability}. For each of these challenges, we develop a corresponding response: compact open-weight transformers that can be run locally, task-specific fine-tuning for termination estimation, imbalance-aware training combined with heterogeneous ensembles, and a syntax-aware attribution pipeline for explanation.

\paragraph{\textbf{Challenge 1: deployable models for privacy-sensitive program analysis.}}
Although frontier LLMs demonstrate strong reasoning capabilities, they are often too large, too costly, or too dependent on external infrastructure for practical use in many software-engineering settings. In industrial and security-sensitive environments, source code frequently cannot be sent to external cloud services because of privacy, compliance, or intellectual-property constraints. Analysis tools must instead execute on developer machines or within on-premise infrastructure. This motivates our focus on \emph{compact open-weight transformers that can be fine-tuned and deployed locally}. Such models are not only more feasible to run on commodity hardware, but are also more amenable to task-specific adaptation than frontier closed models~\cite{DBLP:journals/corr/abs-2409-00088}.

Against this background, a central question of this paper is whether small transformer models can already be effective for program analysis. We investigate this question in the setting of \emph{program termination prediction}, asking whether lightweight architectures are expressive enough to capture structural cues of termination and non-termination directly from source code. To this end, we study a diverse set of compact transformer models---\emph{albert-base} (Google)~\cite{transAlbertBase}, \emph{distilbert-base} (Hugging Face)~\cite{transDistilbert}, \emph{bert-base} (Google)~\cite{transBert}, \emph{language-perceiver} (DeepMind)~\cite{DeepmindPerceiver}, \emph{bart-base} (Facebook)~\cite{FacebookBart}, and \emph{t5-small} (Google)~\cite{GoogleT5}---whose sizes range from approximately 11M to 110M parameters. At this scale, both fine-tuning and inference remain feasible on commodity hardware.

\paragraph{\textbf{Challenge 2: pretrained transformers are not termination analyzers out of the box.}}
Compact pretrained transformers provide useful representations, but their original objectives are designed for natural language understanding and generation rather than reasoning about program execution semantics. Moreover, the task-specific classification layer is randomly initialized and carries no prior knowledge of termination behaviour. As a result, applying these models without adaptation does not simply degrade performance slightly; rather, it yields predictions with little semantic grounding in the target task. Our response is therefore to \emph{fine-tune} these models on the largest publicly available benchmark for this problem, TerGEC~\cite{TerGEC}, allowing them to learn structural regularities associated with both termination and divergence directly from source code.

\paragraph{\textbf{Challenge 3: extreme class imbalance.}}
Even with task-specific fine-tuning, termination prediction faces a severe data challenge: \emph{non-terminating programs are exceptionally rare}. In TerGEC~\cite{TerGEC}, the dataset contains 20{,}057 terminating programs but only 380 non-terminating ones. Under such skew, naive fine-tuning quickly yields degenerate classifiers that effectively learn to predict that \emph{everything terminates}. To counter this, we explore a suite of imbalance-aware training objectives. Class-balanced binary cross-entropy (BCE-effnum) reweights errors according to effective sample size; focal loss concentrates learning on hard examples; and LDAM enlarges the decision margin for the minority class. At the data level, we additionally apply class-aware sampling so that each mini-batch contains non-terminating programs, preventing optimisation from being dominated by the abundant terminating cases.

\paragraph{\textbf{Challenge 4: no single model captures all minority-class signals.}}
Even with imbalance-aware objectives, a standalone transformer remains fundamentally limited in its ability to recover the rare signals of non-termination. Individual models often achieve strong overall discrimination, indicating that they recognise broad structural patterns associated with termination, yet their sensitivity to non-terminating behaviours remains limited. In practice, this produces classifiers that appear strong on global metrics while failing precisely on the rare behaviours that matter most.

This motivates our next step: rather than searching for a single ``best'' transformer, can we exploit diversity across differently optimised models? Distinct training objectives induce different decision boundaries and therefore emphasise different termination cues. By combining transformers trained under heterogeneous objectives, we aim to capture complementary perspectives on non-termination. One model may be especially sensitive to unbounded counters, another to recursion patterns, and a third to missing decrements or ineffective loop updates. \emph{Ensembles leverage this diversity}, increasing the likelihood that at least one component detects the minority signal. 

Indeed, our results show that ensembling transformers trained with different objectives is a particularly effective way to boost termination prediction under extreme imbalance. The best ensemble consistently improves mAP while maintaining strong AUC, indicating that heterogeneous training objectives yield genuinely complementary signals. More broadly, these findings suggest that, in this setting, \emph{diversity in optimisation matters more than scale alone}.

\paragraph{\textbf{Challenge 5: predictions must be interpretable.}}
A further challenge for the practical use of neural termination predictors is interpretability. When a model flags a program as non-terminating, developers and verification engineers need to understand \emph{why} in order to assess and address the underlying risk. Purely predictive models offer little guidance here, which limits their usefulness in verification pipelines. To bridge this gap, we complement our predictive framework with an attribution pipeline that explains model decisions in terms of program structure. Specifically, we map token-level attributions from multiple transformers onto abstract syntax tree (AST) nodes and aggregate them into a unified explanation. This produces syntax-aware visualisations that highlight which constructs most influenced the model's prediction.

\paragraph{\textbf{Our contributions.}}
We make the following contributions:
\begin{itemize}[noitemsep, topsep=0pt, parsep=0pt,leftmargin=*]
    \item We present a practical framework for \emph{transformer-based program termination prediction} using compact open-weight models that can be fine-tuned and deployed locally, making the approach suitable for privacy-sensitive and on-premise program-analysis settings.
    \item We show that small transformers can learn useful termination cues directly from source code.
    \item We show that imbalance-aware training and class-aware sampling improve detection of rare non-terminating programs.
    \item We introduce heterogeneous transformer ensembles that outperform single models.
    \item We provide evidence that, in this setting, model diversity can beat model scale, outperforming graph-based baselines and much larger off-the-shelf LLMs.
    \item We propose a token-to-AST attribution pipeline for syntax-aware explanations.
\end{itemize}

\section{Our Approach}
\label{sec:approach}

\begin{figure*}[h]
    \centering
    \includegraphics[width=0.98\linewidth]{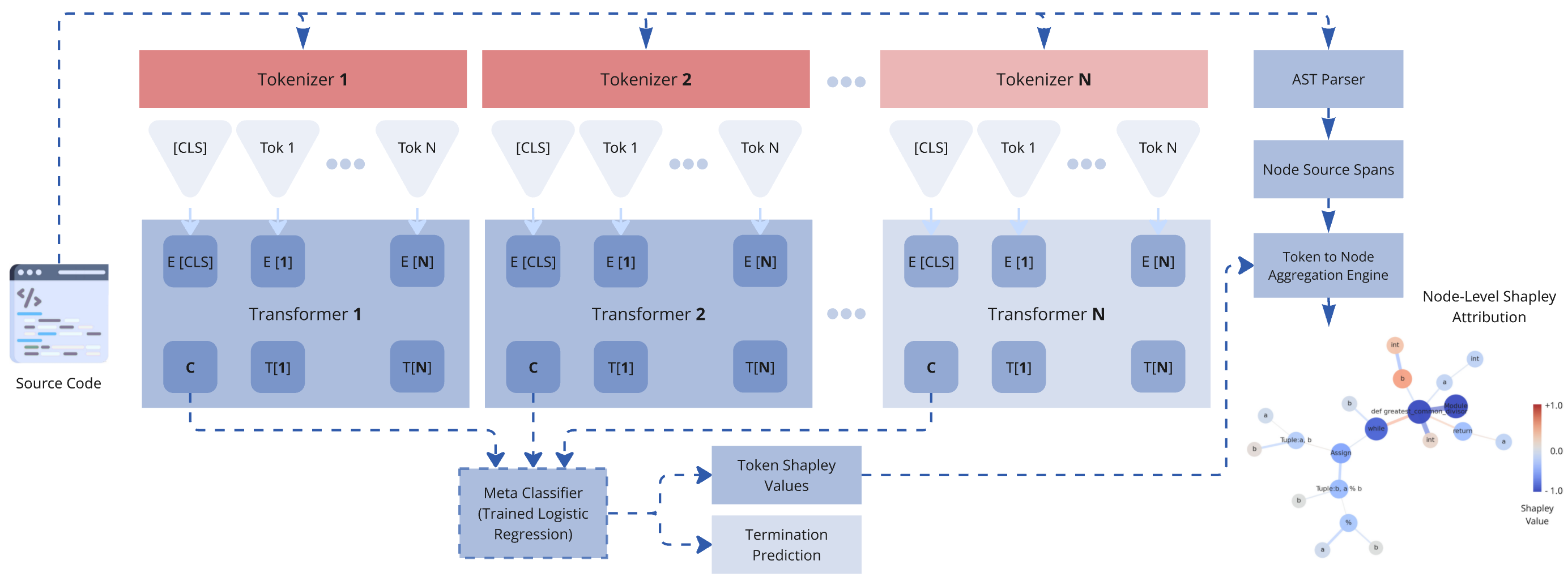}
    \caption{Source code is evaluated by a multi-transformer ensemble using dynamic aggregation to yield a final binary prediction. In parallel, token-level Shapley values are computed and mapped to the source spans of an Abstract Syntax Tree (AST). The resulting alignment produces a structurally attributed AST, localizing the ensemble's predictive attention to specific syntactic nodes.}
    \label{fig:Architecture}
\end{figure*}

Figure~\ref{fig:Architecture} summarizes the overall pipeline. Our method is organised around the same challenges introduced in Section~1. First, to address the challenge of \emph{deployable termination prediction}, we fine-tune a collection of compact pretrained transformers on source code. Second, to address the challenge of \emph{extreme class imbalance}, we train these models under a range of imbalance-aware objectives and sampling strategies. Third, to address the fact that \emph{no single model captures all minority-class signals}, we combine the resulting transformers into an ensemble. Finally, the predictions of this ensemble are paired with a structural attribution pipeline, described in Section~\ref{explain}, which maps token-level explanations back to AST nodes.

Concretely, the input program is first processed by multiple transformer backbones, each with its own tokenizer and classifier head. Their outputs are then combined by an aggregation layer to produce a final binary prediction: terminating or non-terminating. In parallel, we compute token-level Shapley attributions for each transformer and align them with the program's AST, yielding a syntax-aware explanation of the ensemble's decision.

\paragraph{Compact transformers for termination classification.}
To address the deployability and task-adaptation challenges from the introduction, we treat source code as a token sequence and fine-tune compact pretrained transformers for binary classification. Each model receives the program text and produces a contextual representation, which is mapped through a lightweight classification head to a probability of non-termination. This text-first view complements graph-based approaches: instead of constructing ASTs or control-flow graphs as input representations, the model learns directly from source code, allowing us to investigate whether compact transformers can recover structural signals of termination from token-level context alone.

Our study includes a diverse set of compact open-weight backbones with parameter counts ranging from roughly 11M to 110M. Using multiple architectures is important for two reasons. First, it reflects the paper's emphasis on locally deployable models. Second, it introduces architectural diversity, which later becomes useful when constructing ensembles.

\paragraph{Training under extreme class imbalance.}
The main learning challenge is that non-terminating programs are extremely rare. In such settings, naive fine-tuning with standard cross-entropy often produces models that achieve good overall ranking performance while still failing to retrieve the minority class. This is precisely the failure mode highlighted in the introduction: a model may appear successful globally, yet miss the rare non-terminating cases that matter most in practice.

To mitigate this, we train transformers using several imbalance-aware objectives. In addition to the standard cross-entropy baseline, we consider BCE-effnum, Focal loss, and LDAM. These losses address imbalance in complementary ways. BCE-effnum reweights classes according to their effective sample counts, increasing the influence of minority-class errors. Focal loss down-weights easy majority examples so that gradient updates concentrate on hard, often minority-class cases. LDAM reshapes the decision boundary by imposing larger margins for the minority class, making it harder for rare non-terminating programs to be absorbed into the majority region.

At the data level, we further apply class-aware sampling (CAS), ensuring that each mini-batch contains non-terminating programs. This prevents optimisation from being dominated by long stretches of majority-only batches and stabilises learning on the minority class.

\paragraph{From specialised models to ensembles.}
Even with imbalance-aware training, a single transformer captures only part of the minority-class signal. Different objectives and architectures emphasise different cues of non-termination, such as unbounded counter growth, ineffective updates, or recursion patterns. To exploit this diversity, we combine individually fine-tuned transformers into an ensemble.

Our ensemble design mirrors the progression shown in Figure~\ref{fig:Architecture}. Each backbone first produces its own probability estimate for non-termination. These per-model outputs are then passed to an aggregation module, implemented as a lightweight meta-classifier, which learns how to combine the strengths of the component models into a final binary decision. This keeps the aggregation stage simple while still allowing the ensemble to exploit complementary signals across architectures and training objectives.

To study how different forms of diversity affect performance, we construct three ensembles with increasing emphasis on minority-class detection:
\begin{itemize}
    \item \emph{Ensemble~1} combines transformers trained with standard cross-entropy and serves as a performance-oriented baseline;
    \item \emph{Ensemble~2} combines transformers trained with imbalance-aware objectives, testing whether loss-level diversity improves minority detection; and
    \item \emph{Ensemble~3} extends Ensemble~2 with class-aware sampling, combining loss-level and data-level mitigation in a single ensemble.
\end{itemize}
This progression is deliberate. It allows us to separate the effect of objective design from the effect of balanced batch construction, and to test the central hypothesis from the introduction that, in this setting, \emph{model diversity can matter more than model scale}.

\paragraph{Connection to explainability.}
The final component of Figure~\ref{fig:Architecture} addresses the interpretability challenge. In parallel with prediction, we compute token-level Shapley attributions for each transformer, project these attributions back onto source spans, and align them with AST nodes. Aggregating these node-level signals across models yields a structurally attributed AST that highlights which program constructs most influenced the final ensemble decision. Section~\ref{explain} describes this attribution pipeline in detail.

\section{Explainability through Token-to-AST Attribution}
\label{explain}

To investigate the specific code patterns driving the ensemble's decisions and provide actionable feedback for developers, we adopt a game-theoretic attribution framework based on Shapley values ~\cite{EoMShapleyValue} as a principled way to attribute model predictions to input features. Formally, the Shapley score of a feature measures its average marginal contribution to the prediction relative to a baseline. Since exact computation is infeasible, we use standard SHAP-based approximations.

To explain termination predictions at the level of program structure, we extend Shapley attribution from tokens to abstract syntax tree (AST) nodes and aggregate explanations across multiple transformers. The pipeline proceeds as follows:

\paragraph{Step 1: Token-level attribution.}
For each transformer, we compute token-level Shapley values using SHAP. These quantify how individual tokens push the prediction toward termination or non-termination.

\paragraph{Step 2: Mapping tokens to AST nodes.}
Token spans are projected back onto the source code and aligned with AST nodes. Each node receives the sum of Shapley values of its overlapping tokens, producing syntax-aware attribution scores that highlight influential constructs such as loop conditions or recursive calls.

\paragraph{Step 3: Ensemble aggregation.}
Node-level attributions are averaged across models to obtain a unified explanation. This aggregation dampens tokenizer-specific artifacts and emphasizes features that are consistently important across loss functions and transformer variants.

\begin{figure}[h]
    \centering
    \includegraphics[width=0.9\linewidth]{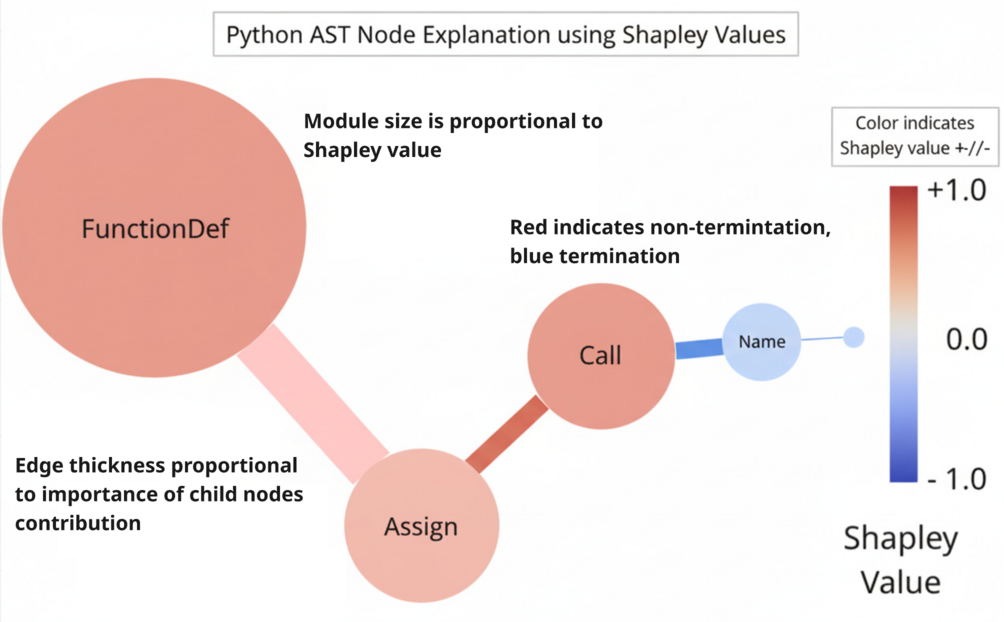}
    \caption{Example attribution graph produced by our pipeline. Larger nodes denote stronger influence; thicker edges emphasize structurally important subtrees.}
    \label{fig:ast-attribution}
\end{figure}

\paragraph{Step 4: Visualization.}
Attributions are rendered as attributed AST graphs: node size reflects attribution magnitude, and edge thickness encodes structural influence (Figure~\ref{fig:ast-attribution}). These visualisations highlight the program regions most responsible for the prediction.

This pipeline moves beyond raw token importances, enabling syntax-aware, ensemble-based explanations of how termination patterns manifest in source code.

\paragraph{Interpreting node-level Shapley scores.}
Let $f(x)$ be the model’s predicted probability of non-termination for a program $x$. For an AST node $n$, its Shapley value $\phi(n)$ measures how much the presence of that node shifts $f(x)$ relative to the baseline. Positive values indicate evidence for non-termination; negative values indicate evidence for termination. Magnitude reflects influence: a larger $|\phi(n)|$ corresponds to a stronger local effect on the decision. Because our models output probabilities, $\phi(n)$ is interpretable in probability points.

Attribution may diffuse through the tree (e.g., between loop predicates and their surrounding blocks). In practice, leaf-level nodes such as comparisons, updates, or recursive calls tend to provide the sharpest signals, while higher-level nodes offer a broader structural view.

\section{Experimental Setup}

All code and data used in this work are available in an anonymized repository: \url{https://anonymous.4open.science/r/TransformerProgramTermination-3040}

\subsection{Dataset and Preprocessing}

We use the TerGEC dataset~\cite{TerGEC}, which provides the largest publicly available benchmark for termination prediction. It consists of Python programs generated by LLMs---specifically, GPT-Neo 125M and GPT-Neo 1.3B---in response to HumanEval~\cite{HumanEval} and MBPP~\cite{MBPP} tasks, labelled as terminating or non-terminating based on execution with a timeout.
The dataset consists of the following subsets:

\noindent$\bullet$ \textit{125M\_}\allowbreak\textit{HumanEval}: 3{,}250 terminating programs and 88 non-terminating;

\noindent$\bullet$ \textit{125M\_}\allowbreak\textit{MBPP}: 1{,}308 terminating programs and 28 non-terminating;

\noindent$\bullet$ \textit{1\_3B\_}\allowbreak\textit{HumanEval}: 4{,}485 terminating programs and 77 non-terminating;

\noindent$\bullet$ \textit{1\_3B\_}\allowbreak\textit{MBPP}: 10{,}856 terminating programs and 139 non-terminating.

In addition to these Python-only datasets, we also evaluate on the Termination Problems Data Base (TPDB)~\cite{TPDB}, a long-established community C benchmarks curated for the annual International Termination Competition~\cite{termination_competition_portal}. TPDB contains a significantly higher proportion of non-terminating examples, with 158 terminating and 48 non-terminating programs. 

We split the datasets based on the 80/20 rule: 80\% of the data is used for training the model, and the remaining 20\% is held back for testing its performance. 
When using class-aware sampling, we ensure that the training and test sets maintain a representative distribution of the minority and majority classes, preventing the model from becoming biased toward the more prevalent terminating programs.

\subsection{Base Transformer Models}


We fine-tune six widely used pretrained transformer models as the backbone for our study: \textit{albert-base-v2} (Google) ~\cite{transAlbertBase}, \textit{distilbert-base-uncased} (Hugging Face) ~\cite{transDistilbert}, \textit{bert-base-uncased} (Google) ~\cite{transBert}, \textit{language-perceiver} (Deepmind) ~\cite{DeepmindPerceiver}, \textit{bart-base} (Facebook) ~\cite{FacebookBart}, and \textit{t5-small} (Google) ~\cite{GoogleT5}. These models represent a diverse range of architectures—including encoders, encoder-decoder frameworks, and cross-attention mechanisms—with parameter counts ranging from approximately 11M to 110M. 

\subsection{Metrics}
\label{sec:metrics}

To evaluate performance, we report two complementary metrics: AUC and mAP.
AUC (Area Under the ROC Curve) measures how well a classifier ranks
terminating and non-terminating programs across all thresholds.
It can be interpreted as the probability that a randomly chosen
non-terminating program receives a higher score than a randomly
chosen terminating program.

Although ROC-AUC is widely used and relatively stable under class
imbalance, it evaluates ranking performance rather than the ability
to retrieve rare positive examples. In highly skewed settings, a
classifier may achieve a high AUC while still exhibiting a low recall
for the minority class.

To better capture minority-class detection, we therefore also report
mean Average Precision (mAP), which summarizes the precision–recall
trade-off and directly reflects how well the model identifies rare
non-terminating programs.

Reporting both metrics provides a complementary view:
AUC reflects overall separability between classes, while mAP
highlights performance on the rare but critical class of
non-terminating programs.

\subsection{Training}

\paragraph{Base transformer training.}
We fine-tuned all transformer backbones using mini-batch gradient descent with AdamW optimization, weight decay regularization, and a maximum of seven epochs. To mitigate overfitting, we applied early stopping with a patience of ten validation checks, halting training if performance did not improve within this window. Importantly, we enabled checkpointing with automatic restoration of the best-performing model, so that the final model corresponds to the epoch with the highest balanced mAP rather than the last epoch. This procedure effectively prevents the model from overfitting to the training data, ensuring that subsequent evaluation is carried out on the most generalizable and robust model state.  

\paragraph{Ensemble training.}
After fine-tuning the individual transformer models, we combined their predictions using
soft voting. Each base model was applied to the training and test sets to produce probability
estimates for the positive class (non-termination). The ensemble prediction was then obtained
by averaging these probabilities across models, ensuring that the final decision reflected the
collective judgment of all components. This strategy allows the ensemble to exploit the
complementary strengths of transformers trained with different imbalance-aware objectives,
while remaining simple, efficient, and robust. 

\section{Experimental Results}\label{sec:experiments}

To better understand the implications of our experiments, we organize the analysis around specific research questions.

\subsection{Evaluation of the individual transformers (RQ1)}
In this section, to answer the research questions about individual transformers, we focus on the results in Table~\ref{tab:ensembles2}, which reports mAP and AUC for Albert, DistilBERT, and BERT-base trained with different loss functions (BCE-effnum, Focal, and LDAM) as well as with class-aware sampling. For now, we disregard the columns corresponding to Ensemble 3.

\begin{table*}[t]
\centering
\small
\begin{adjustbox}{width=1.0\textwidth, center}
\begin{tabular}{llcccccccc}
\toprule
\multirow{2}{*}{\textbf{Dataset}} 
 & \multirow{2}{*}{\textbf{Model}} 
 & \multicolumn{4}{c}{\textbf{mAP (\%)}} 
 & \multicolumn{4}{c}{\textbf{ROC AUC (\%)}} \\
\cmidrule(lr){3-6} \cmidrule(lr){7-10}
 & & BCE-effnum & Focal & LDAM & Ensemble 3 
   & BCE-effnum & Focal & LDAM & Ensemble 3 \\
\midrule

\multirow{6}{*}{125M HumanEval \cite{HumanEval}} 
 & albert-base (Google) \cite{transAlbertBase}            & 72.67 & 75.74 & 63.80 & \multirow{3}{*}{\textbf{85.23}} 
                                                          & 93.71 & 93.31 & 86.77 & \multirow{3}{*}{\textbf{97.47}} \\
 & distilbert-base (Hugging Face) \cite{transDistilbert}  & 78.15 & 78.01 & 81.26 & 
                                                          & 97.19 & 95.72 & 97.20 & \\
 & bert-base (Google) \cite{transBert}                    & 79.39 & 74.01 & 76.18 & 
                                                          & 96.53 & 96.56 & 96.83 & \\
 & language-perceiver (Deepmind) \cite{DeepmindPerceiver} & 67.96 & 61.06 & 74.06 &
                                                          & 88.94 & 83.48 & 93.10 & \\
 & bart-base (Facebook) \cite{FacebookBart}               & 80.34 & 72.14 & 85.29 &
                                                          & 93.03 & 91.91 & 95.81 & \\
 & t5-small (Google) \cite{GoogleT5}                      & 63.79 & 58.97 & 60.97 &
                                                          & 83.63 & 80.10 & 85.09 & \\
\midrule

\multirow{6}{*}{125M MBPP \cite{MBPP}} 
 & albert-base (Google) \cite{transAlbertBase}            & 65.54 & 56.68 & 50.78 & \multirow{3}{*}{\textbf{74.59}} 
                                                          & 69.96 & 78.78 & 58.33 & \multirow{3}{*}{\textbf{98.63}} \\
 & distilbert-base (Hugging Face) \cite{transDistilbert}  & 56.25 & 73.38 & 58.82 & 
                                                          & 68.29 & 92.62 & 84.03 & \\
 & bert-base (Google) \cite{transBert}                    & 53.59 & 54.14 & 52.22 & 
                                                          & 74.98 & 78.25 & 59.32 & \\
 & language-perceiver (Deepmind) \cite{DeepmindPerceiver} & 61.56 & 53.64 & 54.38 &
                                                          & 68.14 & 70.19 & 74.52 & \\
 & bart-base (Facebook) \cite{FacebookBart}               & 54.17 & 53.75 & 71.22 &
                                                          & 86.24 & 83.12 & 66.46 & \\
 & t5-small (Google) \cite{GoogleT5}                      & 49.51 & 51.04 & 53.98 &
                                                          & 40.91 & 53.92 & 57.72 & \\
\midrule

\multirow{6}{*}{1.3B HumanEval \cite{HumanEval}} 
 & albert-base (Google) \cite{transAlbertBase}            & 68.79 & 74.22 & 63.69 & \multirow{3}{*}{75.60} 
                                                          & 97.93 & \textbf{98.11} & 95.90 & \multirow{3}{*}{98.06} \\
 & distilbert-base (Hugging Face) \cite{transDistilbert}  & 62.54 & 66.50 & 70.46 & 
                                                          & 97.05 & 97.44 & 98.08 & \\
 & bert-base (Google) \cite{transBert}                    & 65.71 & \textbf{76.40} & 66.66 & 
                                                          & 96.31 & 98.09 & 96.57 & \\
 & language-perceiver (Deepmind) \cite{DeepmindPerceiver} & 51.30 & 57.30 & 65.29 &
                                                          & 65.81 & 92.35 & 96.73 & \\
 & bart-base (Facebook) \cite{FacebookBart}               & 62.57 & 67.69 & 66.70 &
                                                          & 94.65 & 97.02 & 96.65 & \\
 & t5-small (Google) \cite{GoogleT5}                      & 58.53 & 60.56 & 53.40 &
                                                          & 78.54 & 78.26 & 75.83 & \\
\midrule

\multirow{6}{*}{1.3B MBPP \cite{MBPP}} 
 & albert-base (Google) \cite{transAlbertBase}            & 60.39 & 54.25 & 51.50 & \multirow{3}{*}{\textbf{75.32}} 
                                                          & 94.70 & 85.27 & 75.33 & \multirow{3}{*}{\textbf{96.38}} \\
 & distilbert-base (Hugging Face) \cite{transDistilbert}  & 56.19 & 63.90 & 63.68 & 
                                                          & 93.79 & 94.73 & 95.42 & \\
 & bert-base (Google) \cite{transBert}                    & 63.93 & 62.52 & 62.64 & 
                                                          & 92.72 & 94.62 & 94.86 & \\
 & language-perceiver (Deepmind) \cite{DeepmindPerceiver} & 52.66 & 54.25 & 52.48 &
                                                          & 77.79 & 79.82 & 77.30 & \\
 & bart-base (Facebook) \cite{FacebookBart}               & 61.97 & 58.66 & 56.08 &
                                                          & 94.83 & 93.53 & 87.50 & \\
 & t5-small (Google) \cite{GoogleT5}                      & 56.65 & 58.17 & 58.04 &
                                                          & 78.55 & 80.86 & 87.53 & \\
\midrule

\multirow{6}{*}{TPDB \cite{TPDB}} 
 & albert-base (Google) \cite{transAlbertBase}            & 89.85 & 91.94 & 91.85 & \multirow{3}{*}{\textbf{95.72}} 
                                                          & 95.96 & 93.27 & 92.93 & \multirow{3}{*}{\textbf{97.31}} \\
 & distilbert-base (Hugging Face) \cite{transDistilbert}  & 89.98 & 91.51 & 90.34 & 
                                                          & 91.25 & 91.58 & 90.57 & \\
 & bert-base (Google) \cite{transBert}                    & 89.47 & 91.25 & 94.57 & 
                                                          & 91.58 & 90.24 & 95.62 & \\
 & language-perceiver (Deepmind) \cite{DeepmindPerceiver} & 74.42 & 96.21 & 87.09 &
                                                          & 77.10 & 98.32 & 90.91 & \\
 & bart-base (Facebook) \cite{FacebookBart}               & 76.70 & 90.48 & 94.60 &
                                                          & 81.14 & 94.61 & 97.31 & \\
 & t5-small (Google) \cite{GoogleT5}                      & 89.64 & 88.05 & 95.06 &
                                                          & 90.24 & 91.25 & 96.30 & \\
\bottomrule
\end{tabular}
\end{adjustbox}
\caption{Comparison of transformer models trained with BCE-effnum, Focal, LDAM, and class-aware sampling, alongside the aggregated Ensemble~3.}
\label{tab:ensembles2}
\end{table*}

\begin{tcolorbox}[question]
\textbf{RQ1(a):} Can transformers be trained to effectively estimate termination?
\end{tcolorbox}

Table~\ref{tab:ensembles2} shows that fine-tuned transformer models achieve strong performance on the termination classification task. Most base models obtain \emph{AUC values above 90\%}, demonstrating reliable overall discrimination between terminating and non-terminating programs. 

However, \emph{mAP scores are much lower and more variable}, ranging between 50–82\% depending on the dataset and model. 

The consistent gap between high AUC and considerably lower mAP indicates that, while transformers perform well in overall discrimination, they struggle to reliably detect non-terminating programs, which are severely underrepresented in the datasets. 

\begin{tcolorbox}[answer]
\textbf{RQ1(a) Answer:} Transformers demonstrate effective termination estimation through a high AUC, but their much lower mAP indicates poor performance in detecting minority classes. This discrepancy highlights that the primary challenge in termination prediction is \emph{achieving robust performance on the minority class under conditions of severe imbalance}.
\end{tcolorbox}

\begin{tcolorbox}[question]
\textbf{RQ1(b):} Which class-imbalance–aware loss functions best improve minority-class detection? How about overall discrimination performance?
\end{tcolorbox}

Across datasets in Table~\ref{tab:ensembles2}, \emph{LDAM and Focal loss most often yield the strongest gains in mAP, showing that they are the most effective objectives for improving minority-class detection.} In contrast, AUC values remain consistently high across all loss functions, with only small variations, reflecting that overall discrimination performance is less sensitive to the choice of imbalance-aware objective.

We hypothesize that this difference arises from how the losses shape learning. BCE-effnum reweights errors inversely to class frequency, but this adjustment is coarse: it increases the penalty for minority mistakes without distinguishing between ``easy'' and ``hard'' cases. By contrast, Focal loss dynamically down-weights well-classified examples and concentrates gradient updates on harder, often misclassified non-terminating programs. LDAM goes further by applying larger decision margins to minority classes, explicitly reshaping the classifier’s boundary to carve out more space for rare non-terminating programs. These mechanisms likely explain why LDAM and Focal deliver stronger improvements in mAP, while BCE-effnum provides only modest gains.

\begin{tcolorbox}[answer]
\textbf{RQ1(b) Answer:} LDAM and Focal loss are most effective for improving minority-class detection (mAP), while overall discrimination performance (AUC) remains high across all losses. Their advantage over BCE-effnum likely stems from focusing learning on hard minority cases (Focal) and reshaping decision boundaries to protect minority recall (LDAM).
\end{tcolorbox}

\subsection{Evaluation of the ensembles (RQ2)}

Here, we focus on the columns for Ensemble 3 in Table~\ref{tab:ensembles2}, together with the comparative results for all three ensembles reported in Table~\ref{tab:ensembles}.

\begin{tcolorbox}[question]
\textbf{RQ2(a):} How do the three ensembles compare?
\end{tcolorbox}

Across all benchmarks in Table~\ref{tab:ensembles}, \emph{the three ensembles exhibit a clear progression in capability}. The cross-entropy ensemble provides strong overall discrimination but still struggles to identify rare non-terminating programs. Incorporating imbalance-aware objectives (BCE-effnum, Focal, LDAM) in Ensemble 2 substantially improves mAP, indicating better sensitivity to termination failures, though with small or mixed effects on AUC. Ensemble 3, which additionally uses class-aware sampling, yields the most robust performance: it consistently achieves the highest mAP while preserving or even increasing AUC. This pattern holds across HumanEval, MBPP, and TPDB, demonstrating that diversity in training objectives and exposure to minority examples are both necessary to improve termination detection without sacrificing overall separability.

\begin{tcolorbox}[question]
\textbf{RQ2(a):} Does the best ensemble enhance termination prediction relative to single transformer baselines?
\end{tcolorbox}

Table~\ref{tab:ensembles2} shows that \emph{Ensemble 3 consistently matches or outperforms the strongest single models}. It achieves the best mAP on three out of four benchmarks (85.23\% on 125M HumanEval, 74.59\% on 125M MBPP, and 75.32\% on 1.3B MBPP) and the best AUC on two (98.63\% on 125M MBPP and 96.38\% on 1.3B MBPP). Even when a single base model edges out the ensemble on one metric (e.g., DistilBERT on 1.3B HumanEval with 98.11\% AUC), the ensemble remains competitive and more robust overall.

\begin{tcolorbox}[answer]
\textbf{RQ2(a) Answer:}
The ensemble architecture systematically improves termination prediction, delivering higher or more stable mAP and AUC across datasets compared to any single transformer.
\end{tcolorbox}


\begin{table}[t]
\centering
\footnotesize 
\setlength{\tabcolsep}{3pt} 
\begin{adjustbox}{width=1.0\columnwidth, center}
\begin{tabular}{@{}lcccccc@{}}
\toprule
\multirow{2}{*}{\textbf{Dataset}} & 
\multicolumn{2}{c}{\textbf{Ensemble 1}} & 
\multicolumn{2}{c}{\textbf{Ensemble 2}} & 
\multicolumn{2}{c}{\textbf{Ensemble 3}} \\
\cmidrule(lr){2-3} \cmidrule(lr){4-5} \cmidrule(lr){6-7}
 & mAP & AUC & mAP & AUC & mAP & AUC \\
\midrule
125M HE \cite{HumanEval} & 77.61 & 97.26 & 81.72 & 96.38 & \textbf{85.23} & \textbf{97.47} \\
125M MBPP \cite{MBPP}   & 70.86 & 80.53 & 68.86 & 95.36 & \textbf{74.59} & \textbf{98.63} \\
1.3B HE \cite{HumanEval} & \textbf{76.07} & 97.88 & 69.90 & 97.31 & 75.60 & \textbf{98.06} \\
1.3B MBPP \cite{MBPP}   & 62.06 & 95.54 & 66.57 & 95.72 & \textbf{75.32} & \textbf{96.38} \\
TPDB \cite{TPDB}        & 72.82 & \textbf{97.31} & 92.90 & 93.27 & \textbf{95.72} & \textbf{97.31} \\
\bottomrule
\end{tabular}
\end{adjustbox}
\caption{Compact ensemble performance (\%). Best values per metric/dataset in bold. \textbf{Ensemble types:} (1) aggregates only cross-entropy models, (2) aggregates models trained with imbalance-aware losses, (3) imbalance-aware losses and class-aware sampling.}
\label{tab:ensembles}
\end{table}

\begin{figure*}[h]
    \centering
    \includegraphics[width=0.98\linewidth]{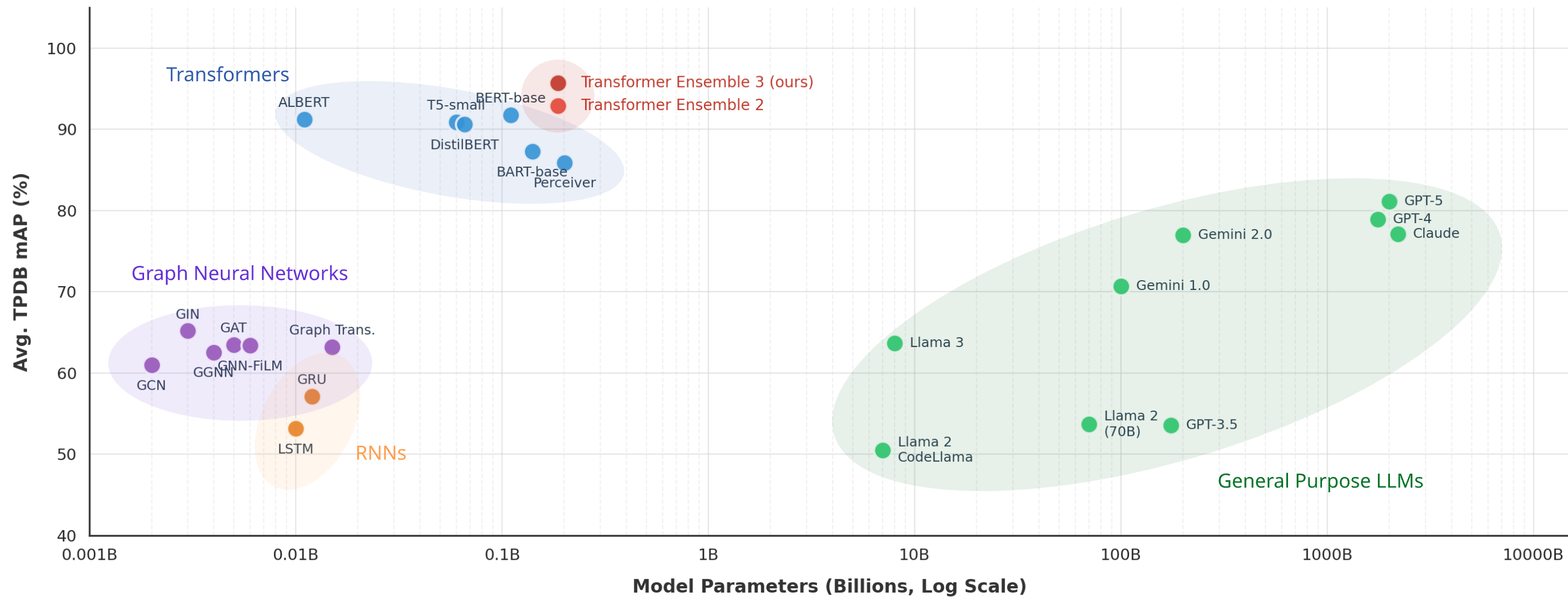}
    \caption{Performance scaling of various model architectures on the TPDB dataset relative to their parameter count (logarithmic scale). For the base transformer models, the mAP reflects the average performance across the BCE-effnum, Focal, and LDAM loss functions to provide a robust baseline. } 
    \label{fig:perforamnce_parameters}
\end{figure*}

\subsection{Comparison to state-of-the-art general-purpose LLMs (RQ3)}

\begin{tcolorbox}[question]
\textbf{RQ3(a)} How does our approach perform relative to state-of-the-art graph neural network-based termination analyzer, TerGEC?
\end{tcolorbox}

Table \ref{tab:gnn_vs_trans} compares our Ensemble 3 with TerGEC \cite{TerGEC}, a graph-enhanced contrastive learning framework that encodes programs as graphs and integrates intra- and inter-class semantics using weighted contrastive and focal losses to address class imbalance. TerGEC represents the current state-of-the-art among graph-based neural approaches.  
Despite this design, our method consistently outperforms TerGEC across all benchmarks. In terms of mAP, which directly reflects minority-class detection, Ensemble~3 achieves much higher scores: 85.23\% vs.\ 70.05\% on 125M HumanEval, 75.60\% vs.\ 61.38\% on 1.3B HumanEval, and 95.72\% vs.\ 68.35\% on TPDB. These margins of 15–27 percentage points confirm that our ensemble is even more effective at detecting non-terminating programs, despite TerGEC's explicit focus on imbalance.  

The improvements in AUC are smaller but consistent (e.g., \ 98.63\% vs.\ 92.32\% on 125M MBPP), reflecting that AUC is less sensitive to imbalance and was already relatively robust for TerGEC. The key observation is that while both approaches aim to address skewed data, \emph{our combination of transformer ensemble and heterogeneous imbalance-aware training strategies yields consistently superior results}. 
We hypothesize that this advantage stems from the richer token-level context captured by transformers, which can model long-range dependencies directly from code. In contrast, GNNs such as TerGEC rely on local message passing over program graphs, and may therefore struggle to capture distant but semantically important relationships without suffering from information bottlenecks like over-squashing. Furthermore, transformers bypass the rigid and often lossy intermediate representations (such as ASTs or CFGs) required by GNNs, allowing self-attention mechanisms to implicitly learn complex control- and data-flow dependencies that heuristic graph constructors might fail to capture.

\begin{tcolorbox}[answer]
\textbf{RQ3(a) Answer:} Although TerGEC was explicitly designed to tackle class imbalance, our Ensemble~3 achieves substantially higher performance across all benchmarks, yielding an absolute improvement in mAP of up to 27\%. We attribute this to transformers' ability to capture long-range token-level context combined with diverse imbalance-aware training, making our ensembles more effective for minority-class detection while also achieving stronger overall discrimination.
\end{tcolorbox}

\begin{table*}[h]
\centering
\begin{adjustbox}{width=1.0\textwidth}
\begin{tabular}{lcccccccccc}
\toprule
\multirow{2}{*}{Model} & 
\multicolumn{2}{c}{125M\_HumanEval \cite{HumanEval}} & 
\multicolumn{2}{c}{125M\_MBPP \cite{MBPP}} & 
\multicolumn{2}{c}{1\_3B\_HumanEval \cite{HumanEval}} & 
\multicolumn{2}{c}{1\_3B\_MBPP \cite{MBPP}} & 
\multicolumn{2}{c}{TPDB \cite{TPDB}} \\
\cmidrule(lr){2-3} \cmidrule(lr){4-5} \cmidrule(lr){6-7} \cmidrule(lr){8-9} \cmidrule(lr){10-11}
 & mAP (\%) & AUC (\%) & mAP (\%) & AUC (\%) & mAP (\%) & AUC (\%) & mAP (\%) & AUC (\%) & mAP (\%) & AUC (\%) \\
\midrule
LSTM \cite{lstm}             & 54.64 & 68.27 & 62.30 & 64.77 & 50.00 & 50.06 & 50.01 & 49.95 & 53.17 & 56.71 \\
GRU  \cite{gru}             & 54.49 & 62.52 & 56.31 & 71.50 & 51.09 & 75.26 & 50.32 & 67.28 & 57.14 & 58.33 \\
\midrule
GCN \cite{GraphConvLayer}              & 60.08 & 81.56 & 71.49 & 85.17 & 58.51 & 88.59 & 58.84 & 90.29 & 61.02 & 84.18 \\
GIN \cite{GIN}              & 63.18 & 86.26 & 66.06 & 85.10 & 60.90 & 90.52 & 59.31 & 91.87 & 65.18 & 87.21 \\
GGNN \cite{GGNN}             & 64.14 & 86.15 & 68.03 & 86.69 & 61.18 & 91.25 & 55.54 & 90.36 & 62.57 & 83.16 \\
GAT \cite{GraphAttention} & 64.51 & 83.54 & 71.54 & 87.22 & 59.33 & 91.23 & 59.61 & 92.08 & 63.50 & 88.89 \\
GNN-FiLM          & 64.92 & 85.21 & 65.90 & 87.22 & 58.67 & 89.56 & 54.71 & 88.74 & 63.38 & 82.49 \\
Graph Transformer \cite{GraphTransformers} & 65.46 & 87.91 & 67.57 & 86.24 & 61.20 & 89.33 & 58.33 & 88.96 & 63.20 & 83.16 \\
\midrule
TerGEC \cite{TerGEC}  & 70.05 & 84.11 & 74.10 & 92.32 & 61.38 & 90.20 & 60.61 & 93.33 & 68.35 & 91.58 \\
\midrule
{\textbf{Ensemble 3 (ours)}} 
              & \textbf{85.23} & \textbf{97.47} & \textbf{74.59} & \textbf{98.63} & \textbf{75.60} & \textbf{98.06} & \textbf{75.32} & \textbf{96.38} & \textbf{95.72} & \textbf{97.31} \\
\bottomrule
\end{tabular}
\end{adjustbox}
\caption{Performance comparison of sequence models, Graph Neural Networks, TerGEC, and our proposed Ensemble 3 across various datasets (HumanEval, MBPP, and TPDB). Evaluation metrics include mAP (\%) and AUC (\%). Best results are highlighted in bold. }
\label{tab:gnn_vs_trans}
\end{table*}

\begin{tcolorbox}[question]
\textbf{RQ3(b)}: How does our method perform relative to off-the-shelf LLMs?
\end{tcolorbox}

To establish a baseline, we evaluate whether general-purpose models can natively resolve the complexities of termination analysis without specialized training. Table~\ref{tab:llm_metrics} compares our Ensemble~3 with several off-the-shelf LLMs on the TPDB dataset. We restricted the evaluation to TPDB to reduce the cost of running inference with these large models and because TPDB is the most challenging benchmark, with a higher proportion of non-terminating programs and greater semantic diversity compared to the synthetic HumanEval and MBPP datasets.  

The results show that Ensemble~3 clearly outperforms the off-the-shelf models. On TPDB, Ensemble~3 achieves an \emph{mAP} of 95.72\% and an \emph{AUC} of 97.31\%, whereas the best-performing proprietary LLM (GPT-5) attains 81.1\% mAP and 86.1\% AUC.

These results highlight that our task-specific approach is far more effective for termination prediction, particularly for the minority class. Ensemble~3's large advantage in mAP (almost 15\% over GPT-5) shows that it is substantially better at detecting non-terminating programs. We hypothesize that this superiority stems from the fact that proprietary LLMs are trained for broad general-purpose capabilities and may only implicitly encode termination behavior, whereas our ensemble is explicitly fine-tuned with imbalance-aware objectives that directly target minority-class recall.  

\begin{tcolorbox}[answer]
\textbf{RQ3(b) Answer:} On the challenging TPDB benchmark, our Ensemble~3 substantially outperforms proprietary LLMs, achieving 95.72\% mAP and 97.31\% AUC compared to GPT-5’s 81.1\% mAP and 86.1\% AUC. This demonstrates that task-specific imbalance-aware training provides far stronger termination prediction, particularly for minority-class detection, than general-purpose LLMs.
\end{tcolorbox}

\begin{table}[h]
\centering
\begin{adjustbox}{width=0.93\columnwidth, center}
\begin{tabular}{lcccc} 
\toprule
\textbf{Model} & \textbf{mAP} & \textbf{AUC} & \textbf{Accuracy} & \textbf{F1} \\
\midrule
GPT-3.5 \cite{gpt-3.5}       & 53.60 & 54.90 & 32.00 & 23.90 \\
GPT-4 \cite{gpt4}         & 78.95 & 82.32 & 70.87 & 78.87 \\
GPT-5 \cite{gpt5}         & 81.10 & 86.10 & 72.80 & 80.10 \\
\midrule
Gemini 1.0 \cite{gemini}    & 70.70 & 78.60 & 72.80 & 80.80 \\
Gemini 2.0 \cite{gemini}    & 77.00 & 83.10 & 72.80 & 80.10 \\
\midrule
Llama 2 \cite{llama2}       & 50.50 & 49.50 & 25.70 &  9.50 \\
Llama 2-70b \cite{llama2}   & 53.70 & 54.90 & 28.60 & 16.00 \\
Llama 3 \cite{llamaLarge}       & 63.70 & 71.00 & 66.00 & 75.20 \\
CodeLlama-S \cite{codellama}   & 58.90 & 60.10 & 77.20 & 86.80 \\
\midrule
Claude Opus 4.6 \cite{claude}   & 77.12 & 82.94 & 83.71 & 82.52 \\
\midrule
\textbf{Ensemble 3 (ours)} & \textbf{95.72} & \textbf{97.31} & \textbf{95.70} & \textbf{94.51} \\
\bottomrule
\end{tabular}
\end{adjustbox}
\caption{Comparison of off-the-shelf LLMs and Ensemble3 on the TPDB dataset \cite{TPDB} for binary termination estimation.}
\label{tab:llm_metrics}
\end{table}

\begin{tcolorbox}[question]
\textbf{RQ3(c)} How do model family and parameter scale influence the accuracy of estimating program termination?
\end{tcolorbox}

In Table \ref{tab:gnn_vs_trans} we see a clear performance difference among the different architectural categories. Recurrent Neural Networks (LSTM and GRU) exhibit the weakest overall performance across all datasets, frequently yielding mAP scores near 50\% on the 1.3B splits, which highlights their limited capacity to capture the complex, long-range dependencies required for termination analysis. Graph Neural Networks, including variants like GIN and GAT, provide a substantial and consistent improvement over RNNs by leveraging explicit code structure, successfully pushing mAP scores into the 60-71\% range and frequently exceeding 85\% in AUC. However, Transformer-based approaches definitively outclass these structural models; while TerGEC sets a strong baseline, our proposed Ensemble 3 dominates every single dataset and metric, achieving state-of-the-art performance with up to 95.7\% mAP and 97.31\% AUC. Taken together with broader scaling trends, this data reinforces that specialized, moderately-sized Transformer architectures vastly outperform both lightweight structural networks and massive, off-the-shelf LLMs, proving that targeted architectural design is paramount for this highly logical task.

An analysis of the trade-off between model size and performance visualized in Figure \ref{fig:perforamnce_parameters} reveals that raw parameter count does not strictly correlate with accuracy in estimating program termination. Lightweight architectures like Graph neural networks and RNNs are highly parameter-efficient but yield only modest predictive results. In contrast, Base Transformers and our proposed ensembles are moderately heavier, yet they capture complex code semantics far more effectively, achieving state-of-the-art accuracy exceeding 90\% mAP. Conversely, off-the-shelf LLMs represent an extreme over-parameterization for this specific task; despite being thousands of times heavier than the base transformers, they consistently underperform them.


\begin{tcolorbox}[answer]
\textbf{RQ3(c) Answer: Our results demonstrate that specifically trained Transformers outperform both structural models like GNNs and RNNs, as well as general-purpose LLMs, achieving superior accuracy with a fraction of the parameters required by the next best-performing large language models.} 
\end{tcolorbox}

\section{Discussion on Explainability}
\label{sec:discussion-explainability}

A central challenge for applying machine learning in program analysis is interpretability: once a model predicts that a program will not terminate, developers and verification engineers need to understand \emph{why}. To this end, we employed the attribution pipeline described in Section~\ref{explain}, which maps token-level Shapley values to abstract syntax tree (AST) nodes and then aggregates them across multiple transformers. The result is a structural visualisation of the program in which node size reflects attribution magnitude and edge thickness highlights the influence of child nodes. Colors indicate the class influence: red pushes predictions toward non-termination, while blue supports termination.

\begin{figure*}[h]
   \centering
   \includegraphics[width=1\linewidth]{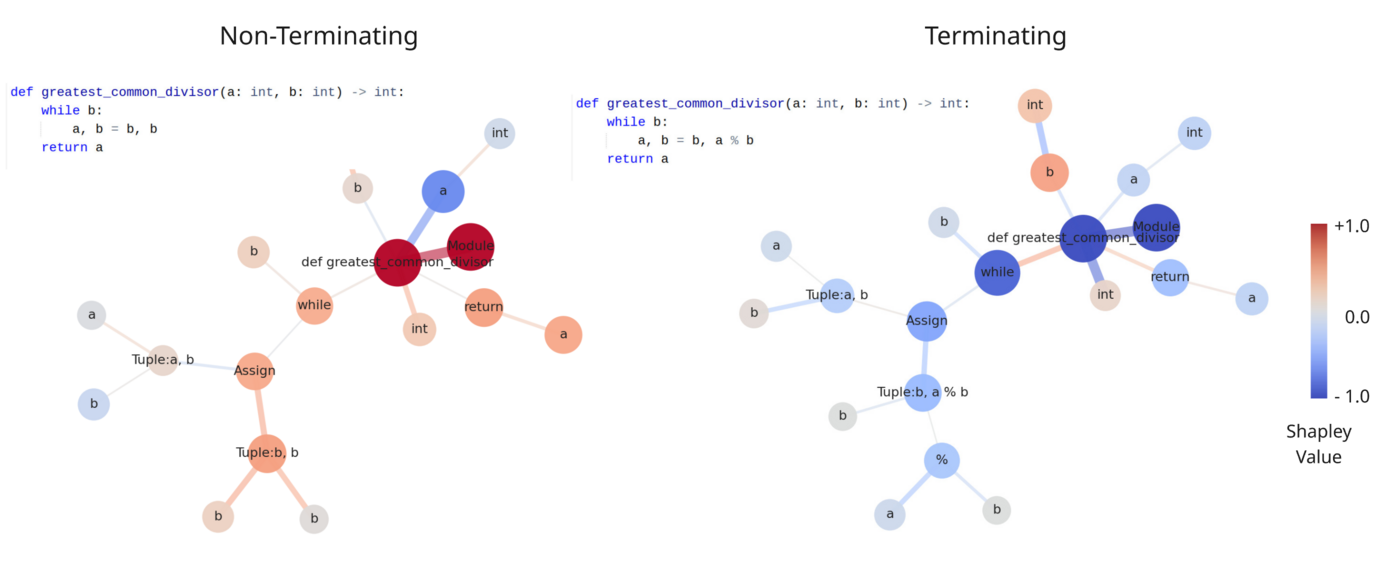}
   \caption[Shapley2]{125M HumanEval example illustrating how attribution isolates the source of non-termination. The original program (right) terminates, while a single modification to the loop condition (left) produces non-termination. The modified loop node and its parent receive the strongest red attribution, clearly identifying the structural cause of the behavioral change.}
   \label{fig:shapley2}
\end{figure*}

\paragraph{Explaining decisions at the AST level.}
Figure~\ref{fig:shapley2} illustrates this process on two closely related programs. The two snippets are identical except for a single loop bound; the original (right) terminates, while the modified version (left) does not. The attribution graph concentrates on the altered loop, where nodes are colored red and shown as the most influential features. This shows that the ensemble explanation is not only faithful to the decision but also isolates the structural cause of the behavioral change.

Manually inspecting many such examples across datasets reveals consistent trends, where (non-)terminating loops and guards are assigned the strongest weights.

\paragraph{Implications.}

This is practically valuable: once a program is flagged as non-terminating, the visualisation can help users pinpoint the problematic region of code. Moreover, ensemble aggregation improves interpretability by filtering out spurious patterns and reinforcing features that are consistently important across diverse models.

\begin{tcolorbox}[answer]
\textbf{Answer:} The ensemble produces faithful and robust explanations by mapping token-level attributions to AST nodes and aggregating them across models. Figure \ref{fig:shapley2} shows an example of these attributions for both a terminating and a non-terminating program, illustrating how modifications to specific nodes directly explain changes in predicted behavior. These visualizations therefore, provide practical diagnostics for identifying high-risk code regions responsible for non-termination.
\end{tcolorbox}

\definecolor{shapPos}{RGB}{181,64,58}   
\definecolor{shapMid}{RGB}{220,219,219} 
\definecolor{shapNeg}{RGB}{76,96,202}   
\section{Related Works}

\subsection{Neural techniques for termination estimation.}

Termination analysis has a long history in program verification. While there are a large number of works on termination analysis, the majority of them employ symbolic reasoning techniques~\cite{DBLP:conf/pldi/CookPR06, DBLP:conf/tacas/CookSZ13, DBLP:conf/esop/DavidKL15, DBLP:conf/pldi/ChenHLLTTZ18}. Although these methods offer strong formal guarantees, they often struggle with real-world complexities, frequently necessitating manually crafted invariants or sophisticated encodings.

Neural Termination Analysis~\cite{neural-termination} introduced one of the earliest data-driven frameworks for termination, where neural networks are trained to approximate ranking functions and then verified with SMT solvers. This hybrid strategy still inherits the limitations of symbolic reasoning, such as dependence on invariants and logical encodings.
Building on purely neural methods, Alon and David proposed to classify program termination directly from code graphs using graph neural networks (GCNs and GATs)~\cite{TerminationGNN}. Their models achieved high accuracy and further enhanced usability through attention and semantic segmentation, enabling localisation of non-termination causes. However, their approach did not explicitly address class imbalance in termination datasets.
Validating this foundational methodology, Liu et al. successfully replicated the approach before advancing it with TerGEC, a graph-enhanced contrastive learning framework~\cite{TerGEC}. By combining intra-class and inter-class learning with weighted contrastive and focal losses, TerGEC explicitly addresses dataset imbalance and achieves state-of-the-art performance on both Python and C benchmarks.

Compared to these works, our method departs from graph-based representations and instead leverages pretrained transformer encoders. This choice avoids the need for explicit graph construction and exploits the transformers' strength in capturing long-range dependencies directly from program text. More importantly, while prior work largely employed a single imbalance-aware objective (e.g., focal or weighted contrastive loss), we introduce loss-level diversity across multiple models, specialising them through  distinct training objectives (BCE-effnum, focal, LDAM). By subsequently combining these heterogeneous models in an ensemble, our approach provides complementary treatment of the minority class (non-termination) and reduces unfair performance disparities across program families.

\subsection{Fairness and bias mitigation}
Bias mitigation has been studied extensively in the broader machine learning literature, typically through pre-processing, in-processing, and post-processing methods. Pre-processing techniques, such as Reweighing \cite{reweighing} and Fair-SMOTE~\cite{fair-smote}, alter training data distributions to balance subgroup representation. In-processing approaches modify the learning procedure, for example, Adversarial Debiasing~\cite{adv-debiasing} or multi-objective optimization as in Fairway~\cite{fairway}. Post-processing methods like Reject Option Classification~\cite{roc} adjust predictions after training to reduce disparities across groups. Toolkits such as IBM’s AIF360~\cite{aif360} consolidate these methods, making it easier to compare bias mitigation strategies.

Ensemble-based approaches have also emerged, typically combining multiple bias mitigation techniques to amplify fairness gains. For example, prior work integrated pre-processing and in-processing methods \cite{fairway}, or combined multiple pre-processing strategies~\cite{fair-smote}. MAAT~\cite{maat} goes further by combining models optimised for different objectives---fairness and performance---showing that objective diversification can improve the fairness–performance trade-off.

Our approach follows the same principle of leveraging diverse optimization goals, but in a different domain and with a different mechanism. Rather than enforcing fairness across demographic groups, we address technical unfairness in program analysis: the skew against non-terminating programs. Unlike pre-processing or data-debugging approaches, which alter the dataset, our method directly enforces fairness at the loss level, using heterogeneous objectives to shape transformer fine-tuning. This positions our contribution as complementary to existing work, extending fairness-inspired ensemble design to tackle imbalance in software verification tasks.

\section{Threats to Validity}

We discuss potential threats to the validity of our study following the standard categories of internal, external, construct, and conclusion validity.

\paragraph{Internal validity.} 
Since we rely exclusively on existing benchmarks, we did not introduce new data collection or labeling steps ourselves. Thus, threats to internal validity primarily concern our experimental pipeline rather than the datasets. Bias could arise from hyperparameter tuning, model checkpointing, or ensemble aggregation. To mitigate this, we applied the same training and evaluation protocol across all models, employed early stopping with automatic checkpoint restoration, and used soft voting for ensemble aggregation rather than more complex meta-learners that risk overfitting. These measures help ensure that observed differences are attributable to training objectives and ensemble composition rather than artifacts of the pipeline.

\paragraph{External validity.}
Our findings are limited by the representativeness of the datasets we use. TPDB and TerGEC are the largest collections of termination datasets available, but they may not fully capture the distribution of termination behaviors in large-scale, real-world systems. In particular, benchmarks based on program generation (e.g., from HumanEval, MBPP) rely on timeouts as proxies for non-termination. These external choices are outside our control but affect the generalizability of our conclusions. Further evaluation on additional languages, industrial codebases, and alternative benchmarks would strengthen external validity. Similarly, our evaluation is confined to three transformer backbones (Albert, DistilBERT, BERT-base). Although these cover a spectrum of model sizes and capacities, results may differ with larger LLMs or alternative architectures. To address this, we compared our ensembles against off-the-shelf LLMs and TerGEC.

\paragraph{Construct validity.}
We focus on two primary evaluation metrics: Area Under the ROC Curve (AUC) and mean Average Precision (mAP). AUC is insensitive to class imbalance and captures overall discrimination, while mAP highlights precision–recall trade-offs under skewed data and is therefore more indicative of minority-class performance. Although this pair of metrics captures complementary aspects of model behavior, they do not exhaust all fairness concerns. In practice, the costs of false positives and false negatives are asymmetric: overlooking a non-terminating program (false negative) is far more harmful than mistakenly flagging a terminating one (false positive). Future work could incorporate explicitly cost-sensitive metrics or task-specific fairness definitions.

\paragraph{Conclusion validity.}
Finally, we note that our conclusions are based on empirical evidence from a finite set of models, datasets, and training objectives. Although our ensembles consistently outperform baselines in both AUC and mAP, some improvements are modest, and the relative effectiveness of different imbalance-aware losses can vary across datasets. Our study does not establish transformers or ensembles as universally superior to GNN-based approaches, but rather demonstrates that they are a competitive and complementary. 

\section{Conclusion}

This work shows that transformers, when fine-tuned with the right training objectives, can serve as effective predictors of program termination. Although individual models struggle with the extreme imbalance between terminating and non-terminating programs, we demonstrate that this challenge can be overcome through targeted methodological adjustments rather than requiring a different underlying architecture. Loss functions that emphasise minority examples, together with class-aware sampling, substantially improve sensitivity to non-termination. Building on this, we introduced ensembles that combine transformers trained under heterogeneous losses, and found that model diversity consistently yields stronger and more reliable predictions than scaling any single model.

Across all benchmarks, including Python datasets and the C based TPDB suite, our best ensemble achieves state-of-the-art performance, outperforming graph-based approaches such as TerGEC and even large off-the-shelf LLMs. Our attribution pipeline further shows that transformer-based predictors ground their decisions in meaningful program structures, e.g., non-terminating loop heads.

Overall, the results establish transformer ensembles as a practical and complementary approach to termination prediction: they offer strong predictive power, improved fairness under extreme imbalance, and explanations that align with program semantics. 

\nocite{LLMsLinear, Thesis}
\bibliographystyle{ACM-Reference-Format}
\bibliography{acmart}   

\appendix

\end{document}